\documentclass[aps,floatfix,nofootinbib,showpacs,onecolumn,superscriptaddress]{revtex4}
\pdfoutput=1

\usepackage{amsmath}
\usepackage{amssymb}
\usepackage{amsthm}
\usepackage{dcolumn}
\usepackage{epsfig}
\usepackage{graphics}
\usepackage{graphicx}
\usepackage{slashed,epsfig}
\usepackage{longtable}
\usepackage{color}

\definecolor{darkgreen}{rgb}{0,0.5,0}
\definecolor{purple}{rgb}{0.5,0,0.5}
\definecolor{nblue}{rgb}{0.0,0.0,0.50}
\definecolor{scarlet}{rgb}{1.0,0.2,0}
\usepackage[colorlinks=true, pdfstartview=FitV, linkcolor=purple, citecolor= purple, urlcolor=blue]{hyperref}

\newcommand{\nn}{\nonumber}
\newcommand{\beq} {\begin{equation}}
\newcommand{\eeq} {\end{equation}}
\newcommand{\beqa} {\begin{eqnarray}}
\newcommand{\eeqa} {\end{eqnarray}}

\newcommand{\ie}{{\it i.e.}}
\newcommand{\eg}{{\it e.g.}}
\newcommand{\cf}{{\it cf.\ }}

\newcommand{\wrt}{{\it wrt.\ }}

\newcommand{\eq}[1]{(\ref{#1})}
\newcommand{\fig}[1]{Fig.~\ref{#1}}

\newcommand{\inv}[1]{\frac{1}{#1}}
\newcommand{\ket}[1]{\vert{#1}\rangle}
\newcommand{\bra}[1]{\langle{#1}\vert}

\newcommand{\acom}[2]{\left\{{#1},{#2}\right\}}

\newcommand{\bs}[1]{\boldsymbol{#1}}

\newcommand{\mM}{\mathcal{M}}

\newcommand{\mA}{\mathcal{A}}

\newcommand{\xv}{{\bs{x}}}

\newcommand{\pv}{{\bs{p}}}
\newcommand{\kv}{{\bs{k}}}
\newcommand{\qv}{{\bs{q}}}
\newcommand{\bv}{{\bs{b}}}
\newcommand{\ev}{{\bs{e}}}

\newcommand{\lv}{{\bs{\ell}}}

\newcommand{\qu}{{\rm q}}
\newcommand{\qb}{{\rm\bar q}}

\newcommand{\halft}{{\textstyle \frac{1}{2}}}

\newcommand{\lsim}{\lesssim}

\begin{document}

{\par\raggedleft \texttt{HIP-2011-01/TH}\par}
{\par\raggedleft {\it Revised April 14, 2011}\par}
\bigskip{}

\title{Measuring transverse shape with virtual photons}

\author{Paul Hoyer and Samu Kurki}
\affiliation{Department of Physics and Helsinki Institute of Physics\\ POB 64, FIN-00014 University of Helsinki, Finland}

\begin{abstract}

A two-dimensional Fourier transform of hadron form factors allows to determine their charge density in transverse space. We show that this method can be applied to any virtual photon induced transition, such as $\gamma^*(q)+N \to \pi N$. Only Fock states that are common to the initial and final states contribute to the amplitudes, which are determined by the overlap of the corresponding light-front wave functions. Their transverse extent may be studied as a function of the final state configuration, allowing qualitatively new insight into strong interaction dynamics. Fourier transforming the cross section (rather than the amplitude) gives the distribution of the transverse distance between the virtual photon interaction vertices in the scattering amplitude and its complex conjugate. While the measurement of parton distributions in longitudinal momentum depends on the leading twist approximation ($-q^2 \to \infty$ limit), all $q^2<0$ values contribute to the Fourier transform, with the transverse resolution increasing with the available range in $q^2$. We illustrate the method using QED amplitudes.

\end{abstract}


\maketitle

\date{\today}

\section{Introduction} 

The photon has a pointlike coupling to quarks, which makes it a valuable probe of hadronic processes at any virtuality $q^2$. The nucleon charge radius is determined from the slope of $eN \to eN$ as $q\to 0$ \cite{Hand:1963zz}. In this limit the nucleon acts as a static source. At higher $q^2$ one expects to be able to map the charge distribution in more detail. However, the quarks are highly relativistic and move as fast as the probe. Hence the time difference between  spatially separated photon interactions cannot be neglected. The standard three-dimensional Fourier transform, which is appropriate for non-relativistic systems such as atoms, assumes an instantaneous photon interaction and is not applicable for tracing quarks in hadrons. 

The precise relation between the spatial distribution of quarks and hadron form factors was long obscure. It was uncovered only recently via a somewhat circuitous route involving deep inelastic scattering, $eN \to eX$ (DIS) \cite{Drell:1969km,Soper:1976jc,Burkardt:2000za,Diehl:2002he,Ralston:2001xs}. In the Bjorken limit ($q^2 \to -\infty$) the DIS cross section is dominated by photon scattering off the same quark in the amplitude and its complex conjugate, with the interaction vertices separated by a light-like distance. The distribution of the quark longitudinal momentum fraction $x$ is therefore most concisely expressed in terms of nucleon wave functions defined at equal Light-Front (LF) time $x^+=t+z,$ \cite{Drell:1969km,Brodsky:1989pv,Brodsky:2000xy}
\beq\label{fq1}
f_{\qu/N}(x,Q^2)  =  \sum_{n} \Bigl[\prod_{i=1}^{n}\int_0^1dx_{i}\int^Q\frac{d^2\kv_{i}}{16\pi^{3}}\sum_{\lambda_i}\Bigr]16\pi^{3}\delta(1-\sum_{i} x_{i})\,\delta^{(2)}(\sum_{i} \kv_{i}) |\psi_{n}^N(x_i,\kv_{i})|^2\,\sum_{k=1}^n e_k^2\, \delta(x_k-x)
\eeq
Here $\psi_{n}^N(x_{i},\kv_{i})$ is the wave function of a nucleon Fock state with $n$ quarks and gluons, all at the same $x^+$ and with each constituent carrying momentum fraction $x_i$ and transverse momentum  $\kv_i$ relative to the nucleon.\footnote{Throughout this paper we use bold symbols to indicate 2-dimensional transverse vectors. Inessential helicities are suppressed.} A Fock state only contributes to $f_{\qu/N}(x,Q^2)$ provided it contains a quark with $x_k=x$. The absolute square of the wave function allows to regard the quark distribution as a probability density.\footnote{This is only approximate \cite{Brodsky:2002ue}, since Coulomb scattering of the quark between the photon vertices is neglected in \eq{fq1}.}

The possibility to access generalized parton distributions (GPDs) in deeply virtual Compton scattering, $eN \to e\gamma N'$, rekindled interest in studying the distribution of quarks also in transverse space (impact parameter) \cite{Burkardt:2000za,Diehl:2002he,Ralston:2001xs}. Fourier transforming the transverse momentum $\pv(N')-\pv(N)$ dependence of the GPD was shown  to give the impact parameter $\bv$ distribution of quarks (after an extrapolation to $p^+(N)=p^+(N')$). The extraction of GPD's from scattering data is demanding, hence our knowledge of quark distributions $q(x,\bv)$ in both longitudinal momentum and impact parameter remains model dependent.

The GPD's reduce to electromagnetic form factors when integrated over the longitudinal momentum fraction, schematically $\int dx\, q(x,\bv) = F(\bv)$. This revealed the desired relation between hadron form factors in momentum space and their charge density in transverse space. Form factors are much easier to measure than GPD's, which allowed to plot the nucleon density distributions without model dependence \cite{Miller:2007uy,Carlson:2007xd}. 
In form factors the momentum transfer to the target equals the photon momentum, $\qv=\pv(N')-\pv(N)$. There is no notion of a ``leading twist'' approximation for form factors. The Fourier transform \wrt $\qv$ gives the charge density of quarks as a function of impact parameter $\bv$. For the Dirac form factor $F_1^N(Q^2)$ of the nucleon,\footnote{We follow the conventions of Ref. \cite{Diehl:2002he}.}
\beqa\label{rho0}
\rho_0(\bv) &\equiv& \int \frac{d^2 \qv}{(2 \pi)^2} \,  
e^{-i \, \qv \cdot \bv} \, \frac{1}{2 P^+}   
\bra{N(p^+, \halft\qv)}\, J^+(0) \,\ket{ 
N(p^+, -\halft\qv)} = \int_0^\infty \frac{d Q}{2 \pi}\, Q \, J_0(b \, Q) F_1^N(Q^2) \nn\\[2mm]
 &=& \sum_{n}\Bigl[\prod_{i=1}^{n}\int_0^1 dx_{i}\int 4\pi d^{2}\bv_{i}\sum_{\lambda_i}\Bigr]\delta(1-\sum_{i} x_{i})\frac{1}{4\pi}\delta^{(2)}(\sum_{i} x_{i}\bv_{i})\, |\psi_{n}^N(x_{i},\bv_{i})|^2\,\sum_{k=1}^n e_{k}\,\delta^{(2)}(\bv-\bv_{k})
\eeqa
Here the $\psi_{n}^N(x_{i},\bv_{i})$ are LF wave functions of a nucleon state $\ket{N(p^+,\bv_N=0)}$ with `plus' momentum $p^+=p^0+p^3$ and transverse `center-of-momentum' $\bv_N=\sum x_i\bv_i=0$. The quarks and gluons in each $n$-parton Fock state have longitudinal momenta $k_i^+=x_ip^+$ and impact parameters $\bv_i$. Only quarks at transverse position $\bv_k = \bv$ contribute to the charge density at $\bv$. The quark distribution in impact parameter \eq{rho0} is analogous and complementary to the standard parton distribution $f_{\qu/N}(x,Q^2)$ \eq{fq1} in longitudinal momentum.\footnote{The form factor has a single photon vertex and hence no Wilson line. The relation \eq{rho0} is exact (up to higher order electromagnetic corrections) insofar as the LF Fock expansion of hadrons is exact (contributions from partons with $x_i=0$ are neglected).}

An expansion similar to \eq{rho0} pertains also for transition form factors measured in $eN \to eN^*$. As we recall below (Section \ref{basic}) the expression remains diagonal in the LF Fock basis: only Fock states that are common to $N$ and $N^*$ contribute. The corresponding wave functions $\psi_n^N,\,\psi_n^{N^*}$ being distinct their product is no longer positive definite. Nevertheless, the impact parameter distribution reflects the transverse size of the transition process, and has been studied using data on several nucleon resonances \cite{Carlson:2007xd,Tiator:2008kd}.

The expression \eq{rho0} for the impact parameter distribution only assumes the general LF Fock expansion of the intial and final hadronic states. Hence it can be applied also to states with several hadrons in the final (and initial) state. This allows to study the transverse size of photon scattering processes as a function of the relative momenta of the final state hadrons. The method can even be applied at the level of cross-sections, thus not requiring a knowledge of the phase of the scattering amplitudes in the Fourier transform. One obtains then the distribution of the transverse distance between the photon vertices in the amplitude and its complex conjugate. Since no leading twist approximation is implied this type of analysis is particularly suitable for data at moderate values of $q^2$, provided only that the contribution of the $J^+$ current can be isolated. The resolution in impact parameter improves with the range of $q^2$ for which data is available.

\section{Basic formalism} \label{basic}  

We first recall \cite{Burkardt:2000za,Diehl:2002he} the basic steps which lead to the expression corresponding to the nucleon density \eq{rho0} for any final state $f$. The lepton scattering amplitude is
\beq\label{mamp}
\mM(\ell N \to \ell' f) = -e^2 \bar u(\ell')\gamma_\mu u(\ell) \inv{q^2} \int d^4 x e^{-iq\cdot x} \bra{f}J^\mu(x)\ket{N(p)}
\eeq
where $q=\ell-\ell'$ is the virtual photon momentum. Using LF spinors \cite{Brodsky:1989pv} quantized along the negative $z$-axis and neglecting the lepton mass,
\beq
\bar u(\ell',\lambda_\ell)\gamma_\mu u(\ell,\lambda_\ell) = \inv{\sqrt{\ell^-(\ell^- -q^-)}}\big[2\ell^-\ell_\mu -\ell^- q_\mu-q^-\ell_\mu+{\bar n}_\mu\ell\cdot q +i\lambda_\ell\epsilon_{\mu\alpha\beta\gamma}{\bar n}^\alpha \ell^\beta q^\gamma \big]
\eeq
where the light-like vector ${\bar n}=(2^+,0^-,\bs{0}_\perp)$ satisfies $\bar n \cdot p=p^- =p^0-p^3$ for any vector $p$. The $J^+$ hadron matrix element dominates in \eq{mamp} in the high energy limit, $\ell^- \to \infty$ at fixed $q$. This limit is also required to formally access all momentum transfers $q$. Hence we consider
\beq\label{mamp2}
\mM(\ell N \to \ell' f) = -e^2\,\frac{\ell^-}{q^2} \bra{f(p_f)}J^+(0)\ket{N(p)}(2\pi)^4 \delta^4(p_f-p-q)\hspace{1cm} (\ell^- \to \infty)
\eeq
The matrix element $\bra{f(p_f)}J^+(0)\ket{N(p)}$ may be viewed as a generalized form factor. Apart from its momentum $p_f= p+q$ there is no restriction on the final hadronic state $f$ which could, \eg, consist of many hadrons (see Section \ref{multihadrons}).

The $J^+$ quark current projects on the $\qu_+$ component of the quark field,
\beq\label{jplus}
J^+(x) = e_q\,\qb(x)\gamma^+\qu(x) = 2e_q\,\qu_+^\dag(x)\qu_+(x)
\eeq
where $\qu_+(x)=\inv{4}\bar{\slashed{n}}\slashed{n}\qu(x)$ and the light-like vector $n=(0^+,2^-,\bs{0}_\perp)$ satisfies $n\cdot p=p^+$ for an arbitrary 4-vector $p$. The $\qu_+$ quark field may be expanded in LF creation and destruction operators at a given LF time. For $x^+=0$,
\beq
\qu_+(0^+,x^-,\xv) =\int\frac{dk^+ d^2\kv}{16\pi^3 k^+}\theta(k^+)\Big[b(k^+,\kv)u_+(k^+)e^{-i\frac{1}{2} k^+x^- +i\kv\cdot \xv}+d^\dag(k^+,\kv)v_+(k^+)e^{i\frac{1}{2}k^+x^- -i\kv\cdot \xv}\Big]
\eeq 
The $u_+$ and $v_+$ spinors are independent of the transverse momentum $\kv$ and normalized according to $u_+^\dag(k^+)u_+(k^+)=k^+$. In terms of operators at a fixed transverse position $\xv$,
\beq
b(k^+,\xv) = \int \frac{d^2\kv}{16\pi^3}\, e^{i\kv\cdot \xv}\, b(k^+,\kv)
\eeq
the quark field is expressed more simply as
\beq\label{qfield}
\qu_+(0^+,x^-,\xv) =\int\frac{dk^+}{k^+}\theta(k^+)\Big[b(k^+,\xv)u_+(k^+)e^{-i\frac{1}{2}k^+x^-}+d^\dag(k^+,\xv)v_+(k^+)e^{i\frac{1}{2}k^+x^-}\Big]
\eeq

The transverse momentum eigenstates may be expanded in impact parameter states 
\beq\label{ptob}
\ket{p^+,\pv}= 4\pi\int d^2\bv\, e^{i\pv\cdot\bv}\ket{p^+,\bv}
\eeq
which have the LF ($x^+=0$) Fock expansion 
\beq\label{fexp}
\ket{p^+,\bv}=\inv{4\pi}\sum_{n}\Bigl[\prod_{i=1}^{n}\int_0^1 \frac{dx_{i}}{\sqrt{x_{i}}}\int 4\pi d^{2}\bv_{i}\Bigr]\delta(1-\sum_{i} x_{i})\delta^{2}(\bv-\sum_{i} x_{i}\bv_{i})\, \psi_{n}(x_{i},\bv_{i}-\bv) \prod^{n} b^\dag(x_i p^+,\bv_i)d^\dag(\ )a^\dag(\ )\ket{0}
\eeq
The $n$ operators in each Fock state create quarks ($b^\dag$), antiquarks ($d^\dag$) and gluons ($a^\dag$) with longitudinal momenta $x_i p^+$ at transverse positions $\bv_i$. The specific advantage of the LF Fock expansion is that a hadron with any longitudinal momentum $p^+$ and transverse position $\bv$ is described by the same  LF wave functions $\psi_{n}(x_{i},\bv_{i}-\bv)$, which depend only on the relative coordinates of the partons.

The quark field \eq{qfield} eliminates an operator $b^\dag(x_k p^+,\bv_k)$ at $\bv_k=\xv$ from the Fock expansion \eq{fexp}, according to the anti-commutation relation 
\beq
\acom{b(k^+,\bv)}{b^\dag({k'}^+,\bv')}=\inv{4\pi}k^+\delta(k^+ -{k'}^+)\,\delta^2(\bv-\bv')
\eeq
Thus, suppressing the contribution of the creation operator $d^\dag(k^+,\xv)$ (see below),
\beqa\label{qfock}
\qu_+(0^+,x^-,\xv)\ket{p^+,\bv}&=&\inv{(4\pi)^2}\sum_{n}\Bigl[\prod_{i=1}^{n}\int_0^1 \frac{dx_{i}}{\sqrt{x_{i}}}\int 4\pi d^{2}\bv_{i}\Bigr]\delta(1-\sum_{i} x_{i})\delta^{2}(\bv-\sum_{i} x_{i}\bv_{i})\, \psi_{n}(x_{i},\bv_{i}-\bv)\nn\\
&& \times \sum_k \Big[(-1)^{P_k}\delta^{2}(\bv_k-\xv)u_+(x_k p^+)e^{-\inv{2}x_k p^+ x^-}\prod_{i\neq k}^{n} b^\dag(x_i p^+,\bv_i)d^\dag(\ )a^\dag(\ )\Big]\ket{0}
\eeqa
where the sign $(-1)^{P_k}$ related to operator ordering will be irrelevant, since according to \eq{jplus} the $J^+$ matrix element in \eq{mamp2} is the overlap of two states of the form \eq{qfock}.

To allow a simple interpretation of the amplitude \eq{mamp2} it is essential to choose a frame where $p_f^+ = p^+$.\footnote{In the case of GPD's this condition implies an extrapolation from the experimentally accessible kinematic region. For form factors it amounts to a choice of frame.} A photon with $q^+=0$ cannot create a $\qu\qb$ pair, causing the matrix element to be diagonal in the number of incoming and outgoing quarks. In fact, the initial and final Fock states are identical. As seen from \eq{qfock} the $J^+(0)$ current interacts with a single quark or antiquark\footnote{Due to the anti-commutation of the $d$-operators the charge $e_k$ in \eq{ffb} has opposite sign for quarks and antiquarks.} at $\bv_k=\bs{0}_\perp$ in $\ket{N}$, and similarly  in $\bra{f}$. The remaining $n-1$ partons in $\ket{N}$ must thus be identical to those in $\bra{f}$. The constraints $\sum_{i} x_{i}=1$ in the initial and final states forces also the momentum fraction $x_k$ of the struck quark to be the same. The ``center of momentum'' constraint $\bv=\sum_{i} x_{i}\bv_{i}$ in \eq{qfock} requires the impact parameters of the initial and final states to be equal, 
\beq\label{gffb}
\inv{2p^+}\bra{f(p^+,\bv_f)}J^+(0)\ket{N(p^+,\bv_N)}\equiv\inv{(4\pi)^2}\delta^{2}(\bv_f-\bv_N)\mA_{fN}(-\bv_N)
\eeq
where, after a shift of integration variables $\bv_i \to \bv_i+\bv_N$,
\beq\label{ffb}
\mA_{fN}(\bv)=\inv{4\pi}\sum_{n}\Bigl[\prod_{i=1}^{n}\int_0^1 dx_{i}\int 4\pi d^{2}\bv_{i}\Bigr]\delta(1-\sum_{i} x_{i})\delta^{2}(\sum_{i} x_{i}\bv_{i})
 {\psi_{n}^{f}}^*(x_{i},\bv_{i})\psi_{n}^{N}(x_{i},\bv_{i})\sum_k e_k\delta^{2}(\bv_k-\bv)
\eeq
This expression for the current matrix element in impact parameter space is central for the applications we consider below. For $f=N$ the positivity of $|\psi_{n}^{N}(x_{i},\bv_{i})|^2$ allows the Fourier transform \eq{rho0} of the elastic (helicity non-flip) form factor to be interpreted as a charge density. Even when the final state differs from the initial one its electro-excitation still proceeds only via Fock components $n$ which are common to both. 

As already indicated in \eq{rho0}, the Fourier transform \wrt $\qv$ of the generalized form factor in \eq{mamp2} should be done in a frame where the nucleon and photon momenta are
\beqa\label{frame}
p&=&(p^+,p^-,-\halft \qv)\nn\\
q&=& (0^+,q^-,\qv)\\
p_f&=&(p^+,p^-+q^-,\halft \qv)\nn
\eeqa
The excitation amplitude in impact parameter space is then, using \eq{ptob} and \eq{gffb},
\beqa\label{fta}
\int\frac{d^2\qv}{(2\pi)^2}e^{-i\qv\cdot\bv}\inv{2p^+}\bra{f(p_f)}J^+(0)\ket{N(p)}=&&\\[2mm]
=\int\frac{d^2\qv}{(2\pi)^2}d^2\bv_N d^2\bv_f &&\hspace{-3mm} e^{-i\qv\cdot(\bv+\inv{2}\bv_N +\inv{2}\bv_f)}\frac{(4\pi)^2}{2p^+}\bra{f(p^+,\bv_f)}J^+(0)\ket{N(p^+,\bv_N)} = \mA_{fN}(\bv) \nn
\eeqa
The expansion \eq{ffb} shows that $\mA_{fN}(\bv)$ gets contributions from LF Fock states that are common to the initial and final states (localized at $\bv_N=\bv_f=0$) which have a quark or antiquark at transverse position $\bv_k=\bv$. The range of $\mA_{fN}(\bv)$ in $\bv$ thus reflects the transverse size of the transition process.

The above analysis has previously been applied to elastic and transition electromagnetic form factors \cite{Miller:2007uy,Carlson:2007xd,Tiator:2008kd}. The Fock expansion \eq{fexp} is, however, completely general and applies also to states $\ket{f}$ that consist of several hadrons. This makes it possible to measure the transverse shape of the hadronic states that contribute to $\gamma^*+i\to f$ transitions, for any states $i$ and $f$.

\section{Two-body final states} \label{multihadrons} 

The momentum $p_f = p+q$ of the final state $f$ varies with $q$ in the Fourier transform \eq{fta}, hence the  $p_f$-dependence of all Fock state wave functions in the expansion of $\ket{f(p_f)}$ must be known. As seen from \eq{fexp} the LF wave functions depend only on the relative coordinates of the constituents, not on the total momentum of the state. Final states $\ket{f}=\ket{h_1,\ldots,h_n}$ consisting of several hadrons may be regarded as a particular type of hadronic state, where we are free to specify the relative momenta of the hadrons, each one of which has its own (non-perturbative) Fock expansion. The multi-hadron Fock amplitudes must conform with the general LF rules to ensure the frame independence of the state $\ket{f}$. 
In this Section we specify the LF Fock expansion and the Fourier transform for a two-body ($\pi N$) state, and illustrate it with a tree-level QED amplitude. The multi-hadron case is considered in Section \ref{cross}, where we discuss the Fourier transform of the cross section.

\subsection{Transverse shape analysis of $\bs{\gamma^* N \to \pi N}$}

The standard LF Fock expansion in transverse momentum space for a single pion is \cite{Brodsky:1989pv,Diehl:2002he}
\beq\label{fexppi}
\ket{\pi(p_1^+,\pv_1)}=16\pi^3\sum_{n}\Bigl[\prod_{i=1}^{n}\int_0^1 \frac{dx_{i}}{\sqrt{x_{i}}}\int \frac{d^{2}\kv_{i}}{16\pi^3}\Bigr]\delta\Big(1-\sum_{i} x_{i}\Big)\delta^{2}\Big(\sum_{i} \kv_{i}\Big)\, \psi_{n}^{\pi}(x_{i},\kv_{i}) \prod_{i=1}^{n} b^\dag(x_i p_1^+,x_i\pv_1+\kv_i)\cdots\ket{0}
\eeq
where $\cdots$ stands for the operators which create the remaining $n-1$ partons of the Fock state. As noted above, the wave functions $\psi_{n}^{\pi}(x_{i},\kv_{i})$ are independent of the pion momentum $p_1$. The `plus' momentum of parton $i$ is $x_i p_1^+$ and its transverse momentum is $x_i\pv_1+\kv_i$. The restrictions on the $x_i$ and $\kv_i$ implied by \eq{fexppi} ensure that the parton momenta sum to the total pion momentum in each Fock state.

For a $\pi N$ state we have then the double expansion
\beqa\label{fexppiN}
\ket{\pi(p_1)N(p_2)}&=&(16\pi^3)^2\sum_{n_\pi,n_N}\Bigl[\prod_{i=1}^{n_\pi}\int_0^1 \frac{dx_{i}}{\sqrt{x_{i}}}\int \frac{d^{2}\kv_{i}}{16\pi^3}\Bigr]
\Bigl[\prod_{j=1}^{n_N}\int_0^1 \frac{dy_{j}}{\sqrt{y_{j}}}\int \frac{d^{2}\lv_{j}}{16\pi^3}\Bigr]
\delta\Big(1-\sum_{i=1}^{n_\pi} x_{i}\Big)\delta\Big(1-\sum_{j=1}^{n_N} y_{j}\Big)\\
&\times&\delta^{2}\Big(\sum_{i=1}^{n_\pi} \kv_{i}\Big)\delta^{2}\Big(\sum_{j=1}^{n_N} \lv_{j}\Big)\, \psi_{n_\pi}^{\pi}(x_{i},\kv_{i}) \psi_{n_N}^{N}(y_{j},\lv_{j}) \prod_{i,j} b^\dag(x_i p_1^+,x_i\pv_1+\kv_i)\,b^\dag(y_j p_2^+,y_j\pv_2+\lv_j)\cdots\ket{0}\nn
\eeqa
which should be transformed into the standard form \eq{fexppi}, where parton momenta refer to the total momentum $p_f=p_1+p_2$ of the state. We parametrize the pion and nucleon momenta in terms of a momentum fraction $x$ and relative transverse momentum $\kv$,
\beq\label{relmom}
\begin{array}{ll}
p_1^+ = xp_f^+$\hspace{2cm}$ & \pv_1=x\pv_f+\kv \\[2mm] p_2^+ = (1-x)p_f^+ & \pv_2=(1-x)\pv_f-\kv
\end{array}
\eeq
where $p_f^+=p^+$ and $\pv_f=\inv{2}\qv$ in the frame \eq{frame}. The momentum fractions of the pion and nucleon constituents \wrt $p^+$ are then $x_i'=x\,x_i$ and $y_j'=(1-x)y_j$, respectively. Using this and integrating over $x$ gives
\beq\label{xred}
\int_0^1 dx\,\delta\Big(1-\sum_{i} x_{i}\Big)\delta\Big(1-\sum_{j} y_{j}\Big) = x(1-x)\delta\Big(1-\sum_{i} x_{i}'-\sum_{j} y_{j}'\Big)
\eeq
where $x=\sum_{i} x_{i}'$ on the {\it rhs}. The transverse momenta of the partons may be expressed as
\beq
\begin{array}{ll}
x_i \pv_1+\kv_i = x_i'\pv_f+\kv_i'$\hspace{1cm}$ & \kv_i'=\kv_i+ \kv\, x_i'/x \\[2mm] 
y_j \pv_2+\lv_j = y_j'\pv_f+\lv_j'$\hspace{1cm}$ & \lv_j'=\lv_j-\kv\,y_j'/(1-x)
\end{array}
\eeq
which gives
\beq
\int\frac{d^{2}\kv}{16\pi^3} (16\pi^3)^2 \delta^2 \Big[\sum_i\Big(\kv_i'-\kv\frac{x_i'}{x}\Big)\Big]  \delta^2 \Big[\sum_j\Big(\lv_j'+\kv\frac{y_j'}{1-x}\Big)\Big]
= 16\pi^3 \delta^2 \Big(\sum_i \kv_i'+\sum_j \lv_j'\Big)
\eeq
For an $\ket{f}=\ket{\pi N}$ state specified in terms of the relative hadron momentum by a wave function $\Psi^f(x,\kv)$ we get
\beqa
\ket{\pi N(p_f^+,\pv_f;\Psi^f)}&\equiv&\int_0^1 \frac{dx}{\sqrt{x(1-x)}} \int\frac{d^2\kv}{16\pi^3}\,\Psi^f(x,\kv) \ket{\pi(p_1)N(p_2)}= \label{pinpt}\\
&=& 16\pi^3\sum_{n_\pi,n_N}
\Big[\prod_{i,j}\int_0^1 \frac{dx_{i}'}{\sqrt{x_{i}'}} \frac{dy_{j}'}{\sqrt{y_{j}'}}
\int \frac{d^{2}\kv_i'}{16\pi^3} \frac{d^{2}\lv_j'}{16\pi^3}\Big] \delta\Big(1-\sum_{i} x_{i}'-\sum_{j} y_{j}'\Big) \delta^2 \Big(\sum_i \kv_i'+\sum_j \lv_j'\Big)\nn\\[2mm]
&\times& x(1-x) \Psi^f(x,\kv) \psi_{n_\pi}^{\pi}\Big(\frac{x_{i}'}{x},\,\kv_{i}'-\frac{x_{i}'}{x}\kv\Big) \psi_{n_N}^{N}\Big(\frac{y_{j}'}{1-x},\,\lv_{j}'+\frac{y_{j}'}{1-x}\kv\Big)\nn\\[2mm]
&\times& \prod_i^{n_\pi}\Big[\inv{\sqrt{x}} b^\dag(x_i' p_f^+,x_i'\pv_f+\kv_i')\cdots\Big] \Big[\prod_j^{n_N}\inv{\sqrt{1-x}}b^\dag(y_j' p_f^+,y_j'\pv_f+\lv_j')\cdots\Big]\ket{0}\label{pinfock}
\eeqa
where $x=\sum_{i} x_{i}'$ and $\kv=\sum_i \kv_i$ on the {\it rhs}. This Fock expansion has the standard LF form, implying that the superposition of $\pi N$ plane wave states should be given by $\Psi^f(x,\kv)$ with $x$ and $\kv$ determined by the relations \eq{relmom} in any frame.\footnote{The Fock expansion \eq{pinfock} is for the non-interacting $\ket{\pi N}_{out}$ state at $x^+ \to \infty$. The ${\psi_n^f}^*$ wave functions in \eq{ffb} describe the state \eq{pinpt} at $x^+=0$, before the pion and nucleon have formed.} The standard normalization condition (suppressing the helicities)
\beq
\bra{\pi N(p'^+,\pv';\Psi^f)}\pi N(p^+,\pv\,;\Psi^f)\rangle = 16\pi^3 p^+ \delta(p^+-p'^+)\delta^2(\pv-\pv')
\eeq
implies
\beq
\int_0^1 dx \int \frac{d^2\kv}{16\pi^3} |\Psi^f(x,\kv)|^2 = 1
\eeq

For the wave function $\Psi^f(x,\kv)$ to preserve the invariant mass of the $\pi N$ state it should have support only at fixed
\beq\label{invmass}
p_f^2=(p_1+p_2)^2 = \frac{m_\pi^2}{x}+\frac{m_N^2}{1-x}+\frac{\kv^2}{x(1-x)}
\eeq
This would be satisfied, \eg, by standard partial wave analyses. A superposition defined by the appropriate spherical harmonics in the rest frame ($\pv_f=0$) determines directly the frame independent wave function $\Psi^f(x,\kv)$. However, it is not necessary to constrain the mass \eq{invmass} to be fixed. States with different mass that are produced at the same $q^+=0$ and $\qv$ will differ \wrt $q^-$, which does not affect the Fourier transform.

It is instructive to express the $\pi N$ states also in impact parameter space, again following the conventions for partonic states. For the state \eq{pinpt}
\beq
\ket{\pi N(p^+,\pv\,;\Psi^f)}=4\pi\int d^2\bv \, e^{i\pv\cdot \bv} \ket{\pi N(p^+,\bv\,;\Psi^f)}
\eeq
where
\beq
\ket{\pi N(p^+,\bv\,;\Psi^f)}=4\pi\int \frac{dx\, d^2\bv_\pi \, d^2\bv_N}{\sqrt{x(1-x)}} \delta^2\big[x\bv_\pi+(1-x)\bv_N\big] \Psi^f(x,\bv_\pi)\, \ket{\pi(xp^+,\bv_\pi+\bv)N((1-x)p^+,\bv_N+\bv)}
\eeq
The hadronic wave functions are related according to
\beq\label{psikb}
\Psi^f(x,\bv_\pi)= \int\frac{d^2\kv}{16\pi^3} \exp\left[\frac{i\kv\cdot\bv_\pi}{1-x}\right]\Psi^f(x,\kv)
\eeq
The normalization condition in impact parameter space is
\beq
4\pi \int dx\, d^2\bv_\pi \, d^2\bv_N \delta^2\big[x\bv_\pi+(1-x)\bv_N\big] |\Psi(x,\bv_\pi)|^2 = 1
\eeq

\subsection{Illustration:  $\bs{\ell N \to \ell' N^*(1440) \to \ell'\pi N}$}

We illustrate our approach by considering the $P11(1440)$ (Roper) resonance contribution to the $\gamma^*N\to\pi N$ amplitude. The charge density given by the $N\to N^*(1440)$ transition form factors was previously determined in Ref. \cite{Tiator:2008kd}. Here we include the resonance decay, $N^* \to \pi N$, with a fixed relative momentum between the pion and the nucleon. Thus the hadronic wave function $\Psi^f(x,\kv)$ in \eq{pinpt} is a $\delta$-function in $x$ and $\kv$, compatible with the constraint \eq{invmass}.

Connecting the $N\stackrel{\gamma^*}{\to}N^*$ matrix element given in Eq. (1) of \cite{Tiator:2008kd} with the $N^*\to \pi N$ decay amplitude using the $N^*$ propagator we get the desired contribution to $\bra{\pi(p_1)N(p_2,\lambda_2)}J^+(0)\ket{N(p,\lambda)}/2p^+$ at $p_f^2=M^2$,
\beqa
\mA^{\pi N}_{\lambda\lambda_2}(\qv;x,\kv)  &=& \frac{1}{2p^+}\sum_{\lambda_f}\bra{\pi(p_1)N(p_2,\lambda_2)}N^*(p_f,\lambda_f)\rangle\inv{M^2-p_f^2-iM\Gamma}\bra{N^*(p_f,\lambda_f)}J^+(0)\ket{N(p,\lambda)} \nn\\
&=& \frac{ig^*}{2p^+}\bar u(p_2,\lambda_2)\,\gamma_5\, \frac{\slashed{p}_f+M}{-iM\Gamma}
\left[F_1^{NN^*}(Q^2)\gamma^+ + F_2^{NN^*}(Q^2)\frac{i\sigma^{+,\nu}q_\nu}{M+m_N}\right]u(p,\lambda)
\eeqa
We evaluate this amplitude in the frame \eq{frame} where $Q^2=-q^2=\qv^2$. The final state momenta $p_1$ and $p_2$ are parametrized as in \eq{relmom} and we use the LF helicity spinors given in \cite{Brodsky:1989pv}. The $N^*$ and nucleon masses are denoted $M$ and $m_N$, respectively, $\Gamma$ is the total width of the $N^*$ and $g^*$ is the $\pi N N^*$ coupling constant. The result for the spin flip and non-flip amplitudes are
\beqa
\mA^{\pi N}_{++}(\qv;x,\kv)  &=& \frac{g^*}{\sqrt{1-x}M\Gamma}\left\{\big[M(1-x)-m_N\big]F_1^{NN^*}(Q^2) - e^{i\phi_q-\phi_k}\frac{k_\perp Q}{M+m_N}F_2^{NN^*}(Q^2) \right\}\nn\\[2mm]
\mA^{\pi N}_{+-}(\qv;x,\kv)  &=& \frac{-g^*}{\sqrt{1-x}M\Gamma}\left\{k_\perp e^{i\phi_k}F_1^{NN^*}(Q^2) +  Q\, e^{i\phi_q}\,\frac{M(1-x)-m_N}{M+m_N}F_2^{NN^*}(Q^2)\right\}
\eeqa
where the transverse momenta are expressed as $\qv=q_\perp(\cos\phi_q,\sin\phi_q)$ and $\kv=k_\perp(\cos\phi_k,\sin\phi_k)$.

Fourier transforming to impact parameter space as in \eq{fta} gives
\beqa\label{ApiNb}
\mA^{\pi N}_{++}(\bv;x,\kv)  &=& \frac{g^*}{\sqrt{1-x}M\Gamma}\left\{\big[M(1-x)-m_N\big]\rho_0^{NN^*}(b) + i k_\perp e^{i\phi_b-\phi_k}\rho_1^{NN^*}(b)\right\} \nn\\[2mm]
\mA^{\pi N}_{+-}(\bv;x,\kv)  &=& \frac{-g^*}{\sqrt{1-x}M\Gamma}\left\{k_\perp e^{i\phi_k}\rho_0^{NN^*}(b) - i \big[M(1-x)-m_N\big] e^{i\phi_b}\rho_1^{NN^*}(b)\right\}
\eeqa
where $\bv=b(\cos\phi_b,\sin\phi_b)$ and the charge density distributions
\beqa
\rho_0^{NN^*}(b) &=& \int_0^\infty \frac{dQ}{2\pi}Q J_0(b Q)F_1^{NN^*}(Q^2)\nn\\
\rho_1^{NN^*}(b) &=& \int_0^\infty \frac{dQ}{2\pi}\frac{Q^2}{M+m_N} J_1(b Q)F_2^{NN^*}(Q^2)
\eeqa
are expressed in terms of the transition form factors similarly as in Ref. \cite{Tiator:2008kd}. The impact parameter amplitudes \eq{ApiNb} have the Fock expansion given in \eq{ffb}.

When one does not assume a specific intermediate $\pi N$ state the $\gamma^* N \to \pi N$ amplitude can be expressed in terms of six invariant amplitudes specified, \eg, in \cite{Pasquini:2007fw}. The LF helicity amplitudes in the frame \eq{frame} can be evaluated in terms of such invariant amplitudes similarly as above, allowing to determine charge densities using existing parametrisations of experimental data.

\subsection{Illustration: $\bs{\ell\mu \to \ell'\mu\gamma}$} \label{qedex}

\begin{figure}[t]
\includegraphics[width=11cm]{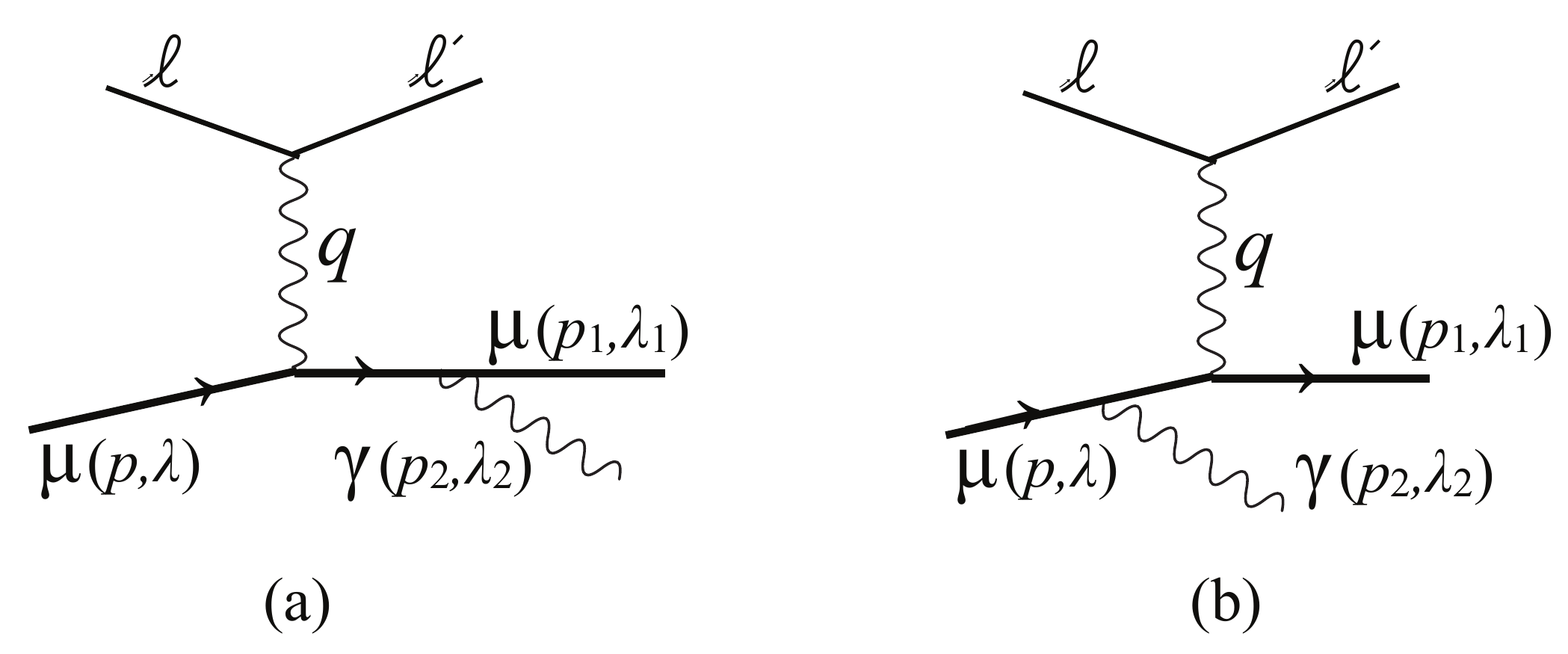}
\caption{\label{Feyn}The two Feynman diagrams contributing to the QED process $\ell\mu \to \ell'\mu\gamma$, when the photon is emitted from the muon.}
\end{figure}

We denote the photon matrix element in the amplitude \eq{mamp2} for $\ell\mu \to \ell'\mu\gamma$ by
\beq\label{muamp}
\mA_{\lambda_1,\lambda_2}^{\mu\gamma}= \inv{2p^{+}}\bra{\mu(p_1,\lambda_1)\gamma(p_2,\lambda_2)}J^+(0)\ket{\mu(p,\lambda=\halft)}
\eeq
where $p+q=p_1+p_2=p_f$ and (as indicated) the initial muon has helicity $\lambda=\halft$. At lowest order, using LF helicity spinors \cite{Brodsky:1989pv} in the frame \eq{frame} and the parametrization \eq{relmom}, diagrams (a) and (b) of \fig{Feyn} give, respectively,
\beq\label{mugamq}
\mA^{\mu\gamma,+\frac{1}{2}}_{+\frac{1}{2}+1}(\qv;x,\kv)  =  2e\sqrt{x}\biggl\{\frac{\ev_{-}\cdot\kv}{(1-x)^{2}m^{2}+\kv^{2}}-\frac{\ev_{-}\cdot[\kv-(1-x)\qv]}{(1-x)^{2}m^{2}+[\kv-(1-x)\qv]^{2}}\biggr\}
\eeq
where $\ev_{\lambda}\cdot\kv=-\lambda e^{i\lambda\phi_{k}}|\kv|/\sqrt{2}$. The corresponding expressions for the other helicity amplitudes are given in the Appendix. The Fourier transform \eq{fta} gives the amplitude for the virtual photon to interact with a muon at impact parameter $\bv$, when the center-of-momentum of the initial and final states is at zero impact parameter:
\beq\label{mugamb}
\mA^{\mu\gamma,+\frac{1}{2}}_{+\frac{1}{2}+1}(\bv;x,\kv)  =  2e\sqrt{x}\biggl[\frac{\ev_{-}\cdot\kv}{(1-x)^{2}m^{2}+\kv^{2}}\delta^{2}(\bv)-\frac{i}{2\sqrt{2}\pi}\frac{m\: e^{-i\phi_{b}}}{1-x}K_{1}(mb)\biggr]\exp\left(-i\frac{\kv\cdot\bv}{1-x}\right)
\eeq
where $\bv=b(\cos\phi_b,\sin\phi_b)$. The first term in \eq{mugamq}
arises from the diagram of \fig{Feyn}(a), 
where the virtual photon vertex is {\it before} the real photon vertex on the muon line, \ie, the exchanged photon interacts with the initial muon. As expected, it contributes to \eq{mugamb} at the initial impact parameter $\bv=0$. The second term is generated by \fig{Feyn}(b), where
the virtual photon interacts with the muon {\it after} the emission of the real photon, and the $\bv$-dependence reflects the impact parameter distribution of the final state muon.

%

The amplitude \eq{mugamb} is in fact given precisely by the overlap \eq{ffb} of the LF Fock amplitudes of the initial $(\mu)$ and final $(\mu\gamma)$ states. According to \eq{muamp} the wave function $\Psi(x,\kv)$ which describes the final state as in \eq{pinpt} is a $\delta$-function in $x$ and $\kv$. In impact parameter space \eq{psikb} gives
\beq\label{psik}
\Psi(x',\bv) =  \delta(x'-x)\sqrt{x(1-x)}\exp\left(i\frac{\kv\cdot\bv}{1-x}\right)
\eeq
The first term in \eq{mugamb} corresponds in \eq{ffb} to the single particle ($n=1$, initial muon) Fock state contribution, which has $x_\mu=1$ and $\bv_\mu=0$. The coefficient of $\delta^{2}(\bv)$ in \eq{mugamb} must therefore be the $c.c.$ of the single muon wave function in the $\mu\gamma$ final state, $\psi^*(\mu\gamma\to\mu)$, which is the same as $\psi(\mu \to \mu\gamma)$ (with reversed sign due to the LF energy denominator). Multiplying the $\mu \to \mu\gamma$ LF wave function given in, \eg, \cite{Brodsky:2000ii},
\beq
-\psi_{+\frac{1}{2}+1}^\uparrow(x,\kv)=\frac{2e}{\sqrt{1-x}}\frac{\ev_{-}\cdot\kv}{(1-x)^{2}m^{2}+\kv^{2}}
\eeq
by $\Psi^*(x',\bv)$ and integrating over $x'$ indeed reproduces the coefficient of $\delta^{2}(\bv)$ in the first term of \eq{mugamb}. The second term in \eq{mugamb} arises from the $n=2$ ($\mu\gamma$) Fock state contribution in \eq{ffb}. It is readily seen to be the product of $4\pi/(1-x)^2$ arising from the integrations in \eq{ffb}, the final state amplitude $\Psi^*(x',\bv)$ and the $\mu \to \mu\gamma$ LF wave function in impact parameter space \cite{Hoyer:2009sg},
\beq\label{lfwfb}
\psi^\uparrow_{+\frac{1}{2}+1}(x,\bv)=-\frac{i}{4\sqrt{2}\pi^2}em\sqrt{1-x}e^{-i\phi_b}K_1(mb)
\eeq

Given the explicit expression for the  QED amplitude \eq{muamp} we may also consider $\mu\gamma$ final states with fixed impact parameter $\bv_\mu'$ of the final muon. Choosing 
\beq
\Psi(x',\bv')=\delta(x'-x)\sqrt{x(1-x)}\:\frac{(1-x)^2}{4\pi}\delta^2(\bv'-\bv_\mu')
\eeq 
we find from \eq{psikb} 
\beq
\Psi(x',\kv)=\delta(x'-x)\sqrt{x(1-x)}\,\exp\Big(-i\frac{\kv\cdot\bv_\mu'}{1-x}\,\Big)
\eeq
Integrating over the relative momentum $\kv$ of the final state with weight $\Psi^*(x',\kv)$ according to \eq{pinpt},
\beqa\label{mugambk}
\mA^{\mu\gamma,+\frac{1}{2}}_{+\frac{1}{2}+1}(\qv;x,\bv_\mu')  &\equiv& \int\frac{d^2\kv}{16\pi^3}\exp\left(i\frac{\kv\cdot\bv_\mu'}{1-x}\right) \mA^{\mu\gamma,+\frac{1}{2}}_{+\frac{1}{2}+1}(\qv;x,\kv) \nn\\[2mm]
& = & -\frac{i}{4\sqrt{2}\pi^{2}}em\sqrt{x}(1-x)\: e^{-i\phi_{b_\mu'}}\: K_{1}\bigl(mb_\mu'\bigr)\biggl[-1+e^{i\qv\cdot\bv_\mu'}\biggr]
\eeqa
The Fourier transform \eq{fta} in $\qv$ then gives
\beq\label{bprimeexpr}
\mA^{\mu\gamma,+\frac{1}{2}}_{+\frac{1}{2}+1}(\bv;x,\bv_\mu') =\sqrt{x(1-x)}\,\psi_{+\frac{1}{2}+1}^{\uparrow}(x,\bv_\mu')\biggl[-\delta^{(2)}(\bv)+\delta^{(2)}(\bv-\bv_\mu')\biggr]
\eeq
Thus the virtual photon interacts either with the initial muon at $\bv=0$ or the final muon at $\bv=\bv_\mu'$. In accordance with \eq{ffb} the distribution is determined by the LF wave function \eq{lfwfb} for $\mu\to\mu\gamma$ (with the sign change noted above).

\section{Cross sections in impact parameter space} \label{cross} 

The superposition \eq{pinpt} and Fourier transform \eq{fta} discussed above require a knowledge of the phase of the scattering amplitude $\bra{f(p_f)}J^+(0)\ket{N(p)}$. Since a partial wave analysis is practical only for a limited subset of all amplitudes it is interesting to ask what information about the transverse structure of the scattering process can be obtained from a Fourier transform of the measured cross section. As we next discuss, this gives the distribution of the transverse distance between the photon interaction vertices in the amplitude and its complex conjugate. 

As in the case of the amplitude \eq{mamp2} we need to isolate the contribution of the $J^+$ current. Here we again consider the high energy limit $s \simeq \ell^- p^+ \to \infty$ at fixed momentum transfer $q=\ell-\ell'$. The Lorentz invariant cross section can then be expressed as
\beq\label{invsig}
\ell^-\frac{d\sigma(\ell N \to \ell' f)}{dq^-\, d^2\qv} \simeq \frac{2\alpha^2}{\pi}\frac{s}{\qv^4}\int d\Pi_f \,\left|\inv{2p^+}\bra{f(p_f)}J^+(0)\ket{N(p)}\right|^2
\eeq
where $d\Pi_f$ is the phase space element of the hadrons in $f$. The frame \eq{frame} can be reached from the $\ell N$ CM by a rotation $\delta\theta \simeq |\qv|/\ell^-$ around the normal to the lepton scattering plane. In the $\ell^- \to \infty$ limit the rotation is infinitesimal and does not affect the finite momentum transfer $q$. Then the Fourier transformation below can be done directly in the $\ell N$ CM.

For a state $f$ with $N_h$ hadrons of momenta $p_i$,
\beq
d\Pi_f(N_h) = \left[\prod_{i=1}^{N_h} \frac{dp_i^+\,d^2\pv_i}{(2\pi)^3 2p_i^+}\right]  (2\pi)^4 \delta^4(p+q-\sum_ip_i)
\eeq
With a LF parametrization as in \eq{relmom},
\beq\label{relmom2}
p_i^+ = x_i p_f^+ \hspace{1cm}  \pv_i=x_i\pv_f+\kv_i \hspace{1cm} (i=1,\ldots,N_h)
\eeq
where $p_f = \sum_i p_i$, we obtain
\beq
d\Pi_f(N_h) = \frac{2(2\pi)^4}{p_f^+}\left[\prod_{i=1}^{N_h} \frac{dx_i\,d^2\kv_i}{(2\pi)^3 2x_i}\right]\delta(1-\sum_i x_i)\,\delta^2(\sum_i \kv_i)\,\delta(p^-+q^- -p_f^- )
\eeq
The initial nucleon $N$ and final state $f$ in the matrix element of \eq{invsig} may be Fourier transformed \eq{ptob} in the frame \eq{frame}, where $\pv_f=-\pv=\halft\qv$ and $q^+=0$. According to \eq{gffb} the matrix element is diagonal in impact parameter. Thus
\beq\label{fsq}
\mathcal{S}_{fN}(\bv) \equiv \int\frac{d^2\qv}{(2\pi)^2}e^{-i\qv\cdot\bv}\,\left|\inv{2p^+}\bra{f(p_f)}J^+(0)\ket{N(p)}\right|^2
=\int d^2\bv_q\, \mA_{fN}(\bv_q)\, \mA_{fN}^*(\bv_q-\bv)
\eeq
Altogether we get for the Fourier transformed cross section,
\beq\label{fts}
\int\frac{d^2\qv}{(2\pi)^2}e^{-i\qv\cdot\bv}\,\qv^4\, \frac{d\sigma(\ell N \to \ell' f)}{d^2\qv}
= (4\pi)^3\alpha^2\sum_{f} \mathcal{S}_{fN}(\bv)
\left[\prod_{i=1}^{N_h}\int \frac{dx_i\,d^2\kv_i}{(2\pi)^3 2x_i}\right]\delta(1-\sum_i x_i)\,\delta^2(\sum_i \kv_i)
\eeq
As indicated, the cross section may include several final states $f$ with different hadron multiplicities $N_h$. 
The amplitudes $\mA_{fN}(\bv_q)$ defined by \eq{fta} can according to \eq{ffb} be expanded in terms of Fock states common to $N$ and $f$. With the initial and final states located at zero impact parameter the struck quark is at impact parameter $\bv_q$. Hence $\mathcal{S}_{fN}(\bv)$ gives the distribution in transverse distance $\bv$ between the quark struck in the amplitude and in its complex conjugate. It has a real part that is even under $\bv \to -\bv$ and an imaginary part that is odd.
A non-vanishing imaginary part requires that the squared matrix element in \eq{fsq} changes when $\qv \to -\qv$. This can be caused by a correlation between $\qv$ and a transverse direction defined by the final state $f$. For example, in \eq{mugamq} the amplitude depends on the angle between $\qv$ and the relative transverse momentum  $\kv$ of the muon and the photon in the final state. A direction can also be specified by a transverse polarization in the initial or final state.

The final phase space integral in \eq{fts} refers to the internal momenta of the final state $f$. {\it E.g.,} in the particular case of $\ket{f}=\ket{\pi(p_1) N(p_2)}$, with $p_1$ and $p_2$ defined by \eq{relmom} and the hadronic wave function $\Psi^f(x,\kv)$ chosen to be a $\delta$-function in $x$ and $\kv$  as in \eq{psik},
\beq\label{pinsig}
\int\frac{d^2\qv}{(2\pi)^2}e^{-i\qv\cdot\bv}\,\qv^4\, \frac{d\sigma(\ell N \to \ell' \pi N)}{d^2\qv\,dx\,d^2\kv} = \frac{\alpha^2}{4\pi^3}\inv{x(1-x)}\, \mathcal{S}_{\pi N,N}(\bv;x,\kv)
\eeq
Thus the impact parameter distribution may be considered for fully exclusive (as well as fully inclusive) cross-sections.

In the case of the $\mu \to \mu \gamma$ example considered in Section \ref{qedex} the impact parameter amplitude $\mA^{\mu\gamma,+\frac{1}{2}}_{+\frac{1}{2}+1}(\bv;x,\kv)$ is given by \eq{mugamb}. The corresponding expression \eq{fsq} for $\mathcal{S}^{\mu\gamma}(\bv;x,\kv)$ is most easily evaluated by directly Fourier transforming $\big|\mA^{\mu\gamma,+\frac{1}{2}}_{+\frac{1}{2}+1}(\qv;x,\kv)\big|^2$,
\beqa\label{sexp}
\mathcal{S}^{\mu\gamma,+\frac{1}{2}}_{+\frac{1}{2}+1}(\bv;x,\kv) = \hspace{15cm} && \\[2mm] 
=4e^{2}x\biggl\{\frac{\kv^{2}/2}{[(1-x)^{2}m^{2}+\kv^{2}]^{2}}\delta^{(2)}(\bv) - im\frac{|\kv|\cos(\phi_{b}-\phi_{k})}{(1-x)^{2}m^{2}+\kv^{2}}\,\frac{K_{1}(mb)}{2\pi(1-x)}
 + \frac{K_{0}(mb)-\halft mb\: K_{1}(mb)}{4\pi(1-x)^{2}}\biggr\}\exp\Big(-i\frac{\kv\cdot\bv}{1-x}\Big) && \nn
\eeqa
The three terms correspond, respectively, to the virtual photon interacting ({\it i}) with the initial muon in both $\mA^{\mu\gamma}$ and $\big(\mA^{\mu\gamma}\big)^*$, ({\it ii}) once with the intial and once with the final muon, and ({\it iii}) twice with the final muon. The imaginary part can be seen to arise from the angular correlation between the lepton scattering plane (defined by $\bv$) and the relative transverse momentum $\kv$ in the final state.
%
\begin{figure}
\includegraphics[width=17cm]{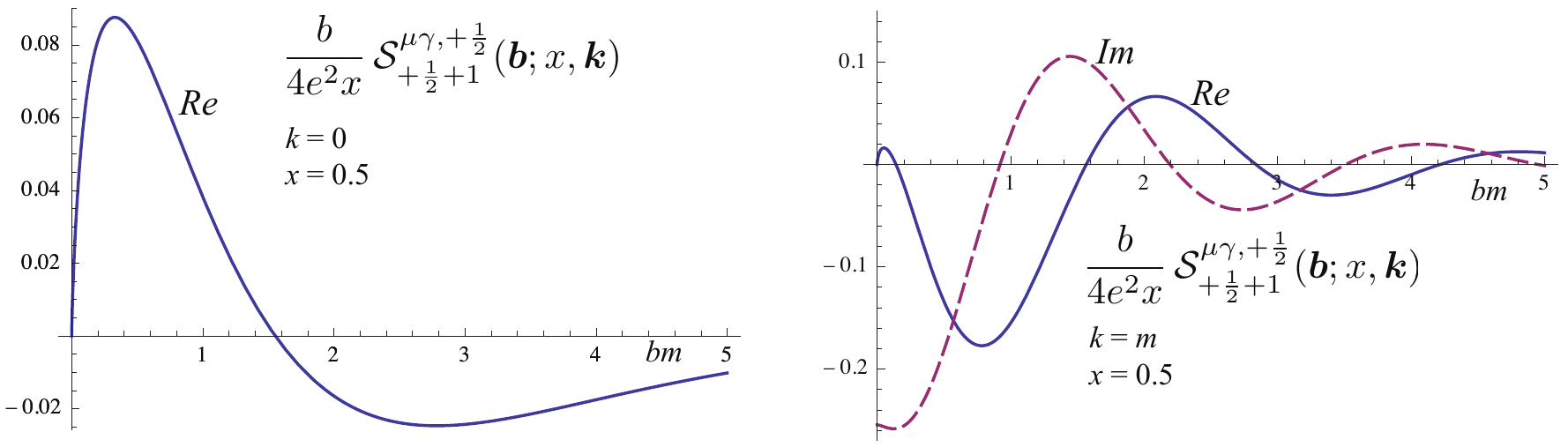}
\caption{\label{Fig1}Plots of the real and imaginary parts of the Fourier transformed cross section $b\,\mathcal{S}^{\mu\gamma,+\frac{1}{2}}_{+\frac{1}{2}+1}(\bv;x,\kv)/4e^2 x$ \eq{sexp} for the QED process $\gamma^*\mu \to \mu\gamma$ with $\kv\parallel\bv$.}
\end{figure}
%
Representative plots of $b\,\mathcal{S}^{\mu\gamma,+\frac{1}{2}}_{+\frac{1}{2}+1}(\bv;x,\kv)/4e^2 x$ are shown in Fig. 1.

\section{Discussion}

The impact parameter analysis of virtual photon induced transition amplitudes and cross-sections appears to open a new window on hadron dynamics. It is complementary to parton distributions in longitudinal momentum, and more economical in using data at all $q^2$, not being restricted to the leading twist ($q^2 \to \infty$) contribution. The analysis can be applied to any final (and initial) state, allowing to study systematic dependencies on, \eg, the mass, relative momenta and flavor content of the state. The $J^+$ component of the electromagnetic current needs to be isolated for a simple Fock state picture.

Only Fock states that are common to the initial and final states contribute to the transition amplitudes \eq{fta}, which are determined by the overlap \eq{ffb} of the corresponding wave functions. This interpretation requires \cite{Burkardt:2000za,Diehl:2002he} a frame like \eq{frame} with $q^+=0$ , where the photon does not create or destroy quark pairs. This is analogous to DIS, where a parton model interpretation is possible only in ``infinite momentum'' frames with $q^+ \leq 0$.

The momentum $p_f=p+q$ of the final state depends on the photon momentum $q$. Relativistic invariance requires that the momenta of all hadrons in $f$ be parametrized as in \eq{relmom2}, with the relative momentum variables $x_i,\kv_i$ being independent of $p_f$. It is possible to form superpositions of final states through weighted integrals over the  $x_i$ and $\kv_i$. In the case of two-particle ($\pi N$) final states we may thus consider states of the form \eq{pinpt} with photon matrix elements
\beq\label{superp}
\bra{\pi N(p_f;\Psi^f)}J^+(0)\ket{N(p)} \equiv \int_0^1 \frac{dx}{\sqrt{x(1-x)}} \int\frac{d^2\kv}{16\pi^3}\,{\Psi^f}^*(x,\kv) \bra{\pi(p_1) N(p_2)}J^+(0)\ket{N(p)}
\eeq
The pion and nucleon momenta are defined by \eq{relmom} and we may freely choose the hadronic wave function $\Psi^f(x,\kv)$. The Fourier transformed amplitudes \eq{fta} get contributions only from quarks at $\bv_q=\bv$, with the initial nucleon and final $\pi N$ states localized at zero impact parameter. The Fourier transform of the squared amplitude \eq{fsq} gives the distribution of the impact parameter difference between the photon interaction vertices in the amplitude and its complex conjugate.

The transverse shape of the contributing Fock states reflects only the distribution of the quarks struck by the photon, not that of the other partons. For example, both compact valence (Brodsky-Lepage \cite{Lepage:1980fj}) Fock states and non-compact (Feynman \cite{Isgur:1984jm,Radyushkin:1984wq}) states may contribute to the elastic form factors of the nucleon at large photon virtualities $|\qv|$. Both types of states will contribute at small $\bv_q$, since the photon interacts only with the $x \to 1$ quark of the Feynman states, whose impact parameter is close to the transverse center-of-momentum $(\bv_N=0)$ of the nucleon.

We expect the impact parameter distribution in $\gamma^* N \to \pi N$ to contract as a function of the relative transverse momentum $\kv$ between the final pion and nucleon. Only compact initial nucleons should have an overlap with $\pi N$ states with high $\kv$, in analogy to the observed color transparency of high energy pions dissociating into exclusive jets with high relative momentum \cite{Aitala:2000hb}.

Large angle photoproduction cross-sections are consistent with constituent counting rules \cite{Brodsky:1973kr,Matveev:1973ra} at surprisingly low energies. Thus $\sigma(\gamma p \to \pi^+ n)$ \cite{Anderson:1976ph} and $\sigma(\gamma p \to K^+ \Lambda)$ \cite{McCracken:2009ra} are both found to be $\propto E_{CM}^{-14}$ at $\theta_{CM}=90^\circ$. Even $\sigma(\gamma D \to p n)$ \cite{Bochna:1998ca} and $\sigma(\gamma\, ^3{\rm He} \to p p(n))$ \cite{Pomerantz:2009sb} obey the rules, scaling as $E_{CM}^{-22}$. The simplest theoretical prediction is based on perturbative QCD, which requires that only transversally compact Fock states contribute at large angles. Data on the $q^2$ dependence of large angle electroproduction would allow to to measure the actual width of the impact parameter distribution.

According to the present analysis all contributing Fock states are common to the initial and final states.
This does not require that heavy quarks $Q$ of final states such as $K\Lambda$ and $D\Lambda_c$ are present in the Fock states of the initial nucleon. Annihilations like $Q\bar Q\,q \to q$ imply that heavy quark final states have Fock components with only light quarks $q$. We would still expect the impact parameter distribution to contract with increasing quark mass, since both the creation and annihilation of heavy quarks has a short transverse range $\sim 1/m_Q$.

The present analysis is also applicable to high energy diffractive processes such as $\gamma^* N \to \rho N$. As may be seen from \eq{invmass} the momentum fraction $x$ of the $\rho$ meson decreases with the CM energy $W\ (W^2=p_f^2)$ as $x \simeq (m_\rho^2+\kv^2)/W^2$. The quarks in the Fock states of the $\rho$ meson have $x_q \lsim x$, and it seems likely that the virtual photon mostly interacts with such low-$x$ quarks. To the extent that the diffractive amplitude has a dominantly imaginary phase and conserves helicity the Fourier analysis may be done at the level of the amplitude, giving the impact parameter distribution of quarks with small $x$.

\acknowledgements

We are grateful for helpful discussions with Stan Brodsky. PH has benefitted from travel support from the Magnus Ehrnrooth foundation. SK acknowledges a PhD study grant from the Jenny and Antti Wihuri Foundation. 

\break

\appendix


\section{}

\vspace{.5cm}
Here we give analytic expressions for the matrix elements of the $J^+$ current in the QED amplitudes $\ell\mu \to \ell'\mu\gamma$ and $\ell\gamma^*_T \to \ell'\mu^+\mu^-$ at lowest order in $\alpha$. They serve to illustrate the general expressions discussed in the text.

\subsection{$\bs{\ell\mu \to \ell'\mu\gamma}$}

We denote the photon matrix element in \eq{muamp} by
\beq
\mA_{\lambda_1,\lambda_2}^{\mu\gamma,\lambda}= \inv{2p^{+}}\bra{\mu(p_1,\lambda_1)\gamma(p_2,\lambda_2)}J^+(0)\ket{\mu(p,\lambda)}
\eeq
where $p+q=p_1+p_2=p_f$ and we parametrize $p_1,p_2$ as in \eq{relmom}. The expressions corresponding to \eq{mugamq} for the various helicity combinations in the frame \eq{frame} are,
\beqa\label{mugamqhel}
\mA^{\mu\gamma,+\frac{1}{2}}_{+\frac{1}{2}+1}(\qv;x,\kv) &=&  2e\sqrt{x}\biggl[\frac{\ev_{-}\cdot\kv}{(1-x)^{2}m^{2}+\kv^{2}}-\frac{\ev_{-}\cdot(\kv-(1-x)\qv)}{(1-x)^{2}m^{2}+(\kv-(1-x)\qv)^{2}}\biggr] \nn\\[2mm]
\mA^{\mu\gamma,+\frac{1}{2}}_{+\frac{1}{2}-1}(\qv;x,\kv) &=&  2e x^{\frac{3}{2}}\biggl[\frac{\ev_{+}\cdot\kv}{(1-x)^{2}m^{2}+\kv^{2}}-\frac{\ev_{+}\cdot(\kv-(1-x)\qv)}{(1-x)^{2}m^{2}+(\kv-(1-x)\qv)^{2}}\biggr]\nn \\[2mm]
\mA^{\mu\gamma,+\frac{1}{2}}_{-\frac{1}{2}+1}(\qv;x,\kv) &=&  \sqrt{2}em\sqrt{x}(1-x)^2\biggl[\frac{1}{(1-x)^{2}m^{2}+\kv^{2}}-\frac{1}{(1-x)^{2}m^{2}+(\kv-(1-x)\qv)^{2}}\biggr]  \nn\\[2mm]
\mA^{\mu\gamma,+\frac{1}{2}}_{-\frac{1}{2}-1}(\qv;x,\kv) &=& 0 \hspace{.5cm} {\rm and} \hspace{.5cm} \mA_{\lambda_1,\lambda_2}^{\mu\gamma,\lambda}(\qv;x,\kv)= \Big[\mA_{-\lambda_1,-\lambda_2}^{\mu\gamma,-\lambda}(-\qv;x,-\kv)\Big]^*
\eeqa
where $m$ is the muon mass and $\ev_{\lambda}\cdot\kv=-\lambda e^{i\lambda\phi_{k}}|\kv|/\sqrt{2}$ with $\kv=|\kv|(\cos\phi_k,\sin\phi_k)$. The Fourier transform \eq{fta} gives
\beqa
\mA^{\mu\gamma,+\frac{1}{2}}_{+\frac{1}{2}+1}(\bv;x,\kv)  &=&  2e\sqrt{x}\biggl[\frac{\ev_{-}\cdot\kv}{(1-x)^{2}m^{2}+\kv^{2}}\delta^{2}(\bv)-\frac{im\: e^{-i\phi_{b}}}{2\sqrt{2}\pi(1-x)} K_{1}(mb)\biggr]\exp\left(-i\frac{\kv\cdot\bv}{1-x}\right) \nn\\[2mm]
\mA^{\mu\gamma,+\frac{1}{2}}_{+\frac{1}{2}-1}(\bv;x,\kv)  &=&  2e x^{\frac{3}{2}}\biggl[\frac{\ev_{+}\cdot\kv}{(1-x)^{2}m^{2}+\kv^{2}}\delta^{2}(\bv)+\frac{im\: e^{+i\phi_{b}}}{2\sqrt{2}\pi(1-x)} K_{1}(mb)\biggr]\exp\left(-i\frac{\kv\cdot\bv}{1-x}\right) \nn\\[2mm]
\mA^{\mu\gamma,+\frac{1}{2}}_{-\frac{1}{2}+1}(\bv;x,\kv)  &=&  \sqrt{2}em\sqrt{x}\biggl[\frac{(1-x)^2}{(1-x)^{2}m^{2}+\kv^{2}}\delta^{2}(\bv)-\frac{1}{2\pi} K_{0}(mb)\biggr]\exp\left(-i\frac{\kv\cdot\bv}{1-x}\right)  \label{mugambhel} \nn \\[2mm]
\mA^{\mu\gamma,+\frac{1}{2}}_{-\frac{1}{2}-1}(\bv;x,\kv)  &=& 0 \hspace{.5cm} {\rm and} \hspace{.5cm} \mA_{\lambda_1,\lambda_2}^{\mu\gamma,\lambda}(\bv;x,\kv)= \Big[\mA_{-\lambda_1,-\lambda_2}^{\mu\gamma,-\lambda}(\bv;x,-\kv)\Big]^*
\eeqa
where $b=|\bv|$. The amplitudes \eq{mugambhel} are given by the overlap \eq{ffb} of LF Fock amplitudes, according to the analysis of \eq{psik}-\eq{lfwfb}. 

The amplitudes for $\mu\gamma$ final states with a fixed impact parameter $\bv_\mu'$ of the final muon are (\cf \eq{mugambk}),
\beqa
\mA^{\mu\gamma,+\frac{1}{2}}_{+\frac{1}{2}+1}(\qv;x,\bv_\mu')  &=&  -\frac{iem}{4\sqrt{2}\pi^{2}}\sqrt{x}(1-x)\: e^{-i\phi_{b_\mu'}}\: K_{1}\bigl(mb_\mu'\bigr)\biggl[-1+e^{i\qv\cdot\bv_\mu'}\biggr] \nn\\[2mm]
\mA^{\mu\gamma,+\frac{1}{2}}_{+\frac{1}{2}-1}(\qv;x,\bv_\mu')  &=&  +\frac{iem}{4\sqrt{2}\pi^{2}} x^\frac{3}{2} (1-x)\: e^{+i\phi_{b_\mu'}}\: K_{1}\bigl(mb_\mu'\bigr)\biggl[-1+e^{i\qv\cdot\bv_\mu'}\biggr] \nn\\[2mm]
\mA^{\mu\gamma,+\frac{1}{2}}_{-\frac{1}{2}+1}(\qv;x,\bv_\mu')  &=&  -\frac{em}{4\sqrt{2}\pi^{2}}\sqrt{x}(1-x)^2\: K_{0}\bigl(mb_\mu'\bigr)\biggl[-1+e^{i\qv\cdot\bv_\mu'}\biggr] \nn \\[2mm]
\mA^{\mu\gamma,+\frac{1}{2}}_{-\frac{1}{2}-1}(\qv;x,\bv_\mu')  &=& 0  \hspace{.5cm} {\rm and} \hspace{.5cm} \mA_{\lambda_1,\lambda_2}^{\mu\gamma,\lambda}(\qv;x,\bv_\mu')= \Big[\mA_{-\lambda_1,-\lambda_2}^{\mu\gamma,-\lambda}(-\qv;x,\bv_\mu')\Big]^*
\eeqa
The Fourier transform \eq{fta} in $\qv$ then gives, for all helicities (\cf \eq{bprimeexpr})
\beq\label{mugambhel2}
\mA^{\mu\gamma,\lambda}_{\lambda_1,\lambda_2}(\bv;x,\bv_\mu') =\sqrt{x(1-x)}\,\psi_{\lambda_1,\lambda_2}^{\mu,\lambda}(x,\bv_\mu')\biggl[-\delta^{(2)}(\bv)+\delta^{(2)}(\bv-\bv_\mu')\biggr]
\eeq
In accordance with \eq{ffb} also the amplitudes \eq{mugambhel2} are thus given by the LF wave functions $\psi^{\mu,\lambda}_{\lambda_1,\lambda_2}(x,\bv'_\mu)$ of the muon in impact parameter space \cite{Hoyer:2009sg}, which describe $\mu^-(p^+,\bv_\mu=0;\lambda) \to \mu^-(xp^+,\bv'_\mu;\lambda_1)\,\gamma((1-x)p^+,-\bv'_\mu\, (1-x)/x;\lambda_2)$. 

The Fourier transformed $\ell\mu\to\ell'\mu\gamma$ cross section \eq{pinsig} is expressed in terms of the squared amplitudes \eq{fsq}, $\mathcal{S}^{\mu\gamma,\lambda}_{\lambda_1\lambda_2}(\bv;x,\kv)=\int d^2\bv_q\, \mA_{\lambda_1,\lambda_2}^{\mu\gamma,\lambda}(\bv_q;x,\kv)\, \mA_{\lambda_1,\lambda_2}^{\mu\gamma,\lambda *}(\bv_q-\bv;x,\kv)$. Their analytical expressions are actually most simply found from the Fourier transform of $|\mA_{\lambda_1,\lambda_2}^{\mu\gamma,\lambda}(\qv;x,\kv)|^2$, 
\beqa
\mathcal{S}^{\mu\gamma,+\frac{1}{2}}_{+\frac{1}{2}+1}(\bv;x,\kv) &=&  4e^{2}x\biggl\{\frac{\kv^{2}/2}{[(1-x)^{2}m^{2}+\kv^{2}]^{2}}\delta^{(2)}(\bv) -\frac{im}{2\pi(1-x)} \frac{|\kv|\cos(\phi_{b}-\phi_{k})}{(1-x)^{2}m^{2}+\kv^{2}}K_{1}(mb) \nn\\[2mm]
 &+& \frac{1}{4\pi(1-x)^2}\Bigl[K_{0}(mb)-\halft mb\: K_{1}(mb)\Bigr]\biggr\}\exp\left(-i\frac{\kv\cdot\bv}{1-x}\right)\nn\\[2mm]
\mathcal{S}^{\mu\gamma,+\frac{1}{2}}_{+\frac{1}{2}-1}(\bv;x,\kv) &=&  x^{2}\; \mathcal{S}^{\mu\gamma,+\frac{1}{2}}_{+\frac{1}{2}+1}(\bv;x,\kv) \nn\\[2mm]
\mathcal{S}^{\mu\gamma,+\frac{1}{2}}_{-\frac{1}{2}+1}(\bv;x,\kv) &=&  2e^{2}m^{2} x\biggl\{\frac{(1-x)^{4}}{[(1-x)^{2}m^{2}+\kv^{2}]^{2}}\delta^{(2)}(\bv) - \frac{1}{\pi}\frac{(1-x)^{2}}{(1-x)^{2}m^{2}+\kv^{2}}K_{0}(mb) \nn\\[2mm]
 &+& \frac{b}{4\pi m}K_{1}(mb)\biggr\} \exp\left(-i\frac{\kv\cdot\bv}{1-x}\right)\label{mugamsf}\nn\\[2mm]
\mathcal{S}^{\mu\gamma,+\frac{1}{2}}_{-\frac{1}{2}-1}(\bv;x,\kv) &=& 0 \hspace{.5cm} {\rm and} \hspace{.5cm} \mathcal{S}_{\lambda_1,\lambda_2}^{\mu\gamma,\lambda}(\bv;x,\kv)= \mathcal{S}_{-\lambda_1,-\lambda_2}^{\mu\gamma,-\lambda}(\bv;x,\kv)
\eeqa

\subsection{$\bs{\ell\gamma^* \to \ell'\mu^-\mu^+}$}

We denote the photon matrix element in \eq{muamp} for $\ell\gamma^* \to \ell'\mu^-\mu^+$ by
\beq
\mA_{\lambda_1,\lambda_2}^{\mu\mu,\lambda}= \inv{2p^{+}}\bra{\mu^-(p_1,\lambda_1)\mu^+(p_2,\lambda_2)}J^+(0)\ket{\gamma^*(p,\lambda)}
\eeq
where $p+q=p_1+p_2=p_f$. The initial photon has virtuality $p^{2}$ and helicity $\lambda=\pm 1$. At lowest order, in the frame \eq{frame} with $p_1,p_2$ parametrized as in \eq{relmom},
\beqa
\mA^{\mu\mu,+1}_{+\frac{1}{2}+\frac{1}{2}}(\qv;x,\kv) &=&  \sqrt{2}em\sqrt{x(1-x)}\biggl[\frac{1}{M^{2}+(\kv-(1-x)\qv)^{2}}-\frac{1}{M^{2}+(\kv +x\qv)^{2}}\biggr] \nn\\[2mm]
\mA^{\mu\mu,+1}_{+\frac{1}{2}-\frac{1}{2}}(\qv;x,\kv) &=&  -2e x^{\frac{3}{2}}\sqrt{1-x} \biggl[\frac{\ev_{+}\cdot(\kv-(1-x)\qv)}{M^{2}+(\kv-(1-x)\qv)^{2}}-\frac{\ev_{+}\cdot(\kv +x\qv)}{M^{2}+(\kv +x\qv)^{2}}\biggr] \nn\\[2mm]
\mA^{\mu\mu,+1}_{-\frac{1}{2}+\frac{1}{2}}(\qv;x,\kv) &=&  -\, \frac{1-x}{x} \mA^{\mu\mu,+1}_{+\frac{1}{2}-\frac{1}{2}}(\qv;x,\kv) \label{mumuqhel}\nn \\[2mm]
\mA^{\mu\mu,+1}_{-\frac{1}{2}-\frac{1}{2}}(\qv;x,\kv) &=& 0 \hspace{.5cm} {\rm and} \hspace{.5cm} \mA_{\lambda_1,\lambda_2}^{\mu\mu,\lambda}(\qv;x,\kv)= \Big[\mA_{-\lambda_1,-\lambda_2}^{\mu\mu,-\lambda}(-\qv;x,-\kv)\Big]^*
\eeqa
where $M^{2}=m^2-x(1-x)p^2$ and $\ev_{\lambda}\cdot\kv=-\lambda e^{i\lambda\phi_{k}}|\kv|/\sqrt{2}$. The  interaction of the virtual photon with the  $\mu^-$ is given by the first terms in \eq{mumuqhel}, and the interaction with the $\mu^+$ by the second term. The Fourier transform \eq{fta} gives
\beqa
\mA^{\mu\mu,+1}_{+\frac{1}{2}+\frac{1}{2}}(\bv;x,\kv) &=& \frac{em}{\sqrt{2}\pi}\sqrt{x(1-x)}\biggl[\frac{K_0\bigl(\frac{M \: b}{1-x}\bigr)}{(1-x)^2} \exp\left(-i\frac{\kv\cdot\bv}{1-x}\right)-\frac{K_0\bigl(\frac{M \: b}{x}\bigr)}{x^2} \exp\left(+i\frac{\kv\cdot\bv}{x}\right)\biggr] \nn\\[2mm]
\mA^{\mu\mu,+1}_{+\frac{1}{2}-\frac{1}{2}}(\bv;x,\kv) &=& \frac{+ieM}{\sqrt{2}\pi}x^{\frac{3}{2}}\sqrt{1-x} \: e^{+i\phi_{b}}\: \biggl[\frac{K_1\bigl(\frac{M \: b}{1-x}\bigr)}{(1-x)^2} \exp\left(-i\frac{\kv\cdot\bv}{1-x}\right)+\frac{K_1\bigl(\frac{M \: b}{x}\bigr)}{x^2} \exp\left(+i\frac{\kv\cdot\bv}{x}\right)\biggr] \nn\\[2mm]
\mA^{\mu\mu,+1}_{-\frac{1}{2}+\frac{1}{2}}(\bv;x,\kv) &=& -\,\frac{1-x}{x} \mA^{\mu\mu,+1}_{+\frac{1}{2}-\frac{1}{2}}(\bv;x,\kv)\nn \\[2mm]
\mA^{\mu\mu,+1}_{-\frac{1}{2}-\frac{1}{2}}(\bv;x,\kv) &=& 0 \hspace{.5cm} {\rm and} \hspace{.5cm} \mA_{\lambda_1,\lambda_2}^{\mu\mu,\lambda}(\bv;x,\kv)= \Big[\mA_{-\lambda_1,-\lambda_2}^{\mu\mu,-\lambda}(\bv;x,-\kv)\Big]^* 
\eeqa
where $b=|\bv|$. The amplitudes for $\mu^-\mu^+$ final states with a fixed impact parameter $\bv_\mu'$ of the final $\mu^-$ (\cf \eq{mugambk} are
\beqa
\mA^{\mu\mu,+1}_{+\frac{1}{2}+\frac{1}{2}}(\qv;x,\bv_\mu')  &=&  \frac{em}{4\sqrt{2}\pi^{2}}\sqrt{x(1-x)} K_0\bigl(\frac{M \: b_\mu'}{1-x}\bigr) \biggl[\exp \biggl(+i\qv\cdot\bv_\mu'\biggr) -\exp \biggl(-i\frac{x \qv\cdot\bv_\mu'}{1-x}\biggr)\biggr] \nn\\[2mm]
\mA^{\mu\mu,+1}_{+\frac{1}{2}-\frac{1}{2}}(\qv;x,\bv_\mu')  &=&  \frac{+ieM}{4\sqrt{2}\pi^{2}} x^{\frac{3}{2}}\sqrt{1-x}\: e^{+i\phi_{b_\mu'}}\: K_1\bigl(\frac{M \: b_\mu'}{1-x}\bigr) \biggl[\exp \biggl(+i\qv\cdot\bv_\mu'\biggr) -\exp \biggl(-i\frac{x \qv\cdot\bv_\mu'}{1-x}\biggr)\biggr] \nn\\[2mm]
\mA^{\mu\mu,+1}_{-\frac{1}{2}+\frac{1}{2}}(\qv;x,\bv_\mu')  &=&  -\,\frac{x}{1-x} \mA^{\mu\mu,+1}_{+\frac{1}{2}-\frac{1}{2}}(\qv;x,\bv_\mu')\nn \\[2mm]
\mA^{\mu\mu,+1}_{-\frac{1}{2}-\frac{1}{2}}(\qv;x,\bv_\mu')  &=& 0  \hspace{.5cm} {\rm and} \hspace{.5cm} \mA_{\lambda_1,\lambda_2}^{\mu\mu,\lambda}(\qv;x,\bv_\mu')= \Big[\mA_{-\lambda_1,-\lambda_2}^{\mu\mu,-\lambda}(-\qv;x,\bv_\mu')\Big]^*
\eeqa
The Fourier transform \eq{fta} in $\qv$ then gives
\beq\label{mumubhel2}
\mA^{\mu\mu,\lambda}_{\lambda_1,\lambda_2}(\bv;x,\bv_\mu') =\sqrt{x(1-x)}\,\psi_{\lambda_1,\lambda_2}^{\gamma,\lambda}(x,\bv_\mu')\biggl[\delta^{(2)}(\bv-\bv_\mu')-\delta^{(2)}(\bv + \frac{x}{1-x}\bv_\mu')\biggr]
\eeq
In accordance with \eq{ffb}, the amplitudes \eq{mumubhel2} are given by the LF wave functions $\psi^{\gamma,\lambda}_{\lambda_1,\lambda_2}(x,\bv'_\mu)$ of the photon in impact parameter space, describing $\gamma^*(p^+,\bv_\gamma=0;\lambda) \to \mu^-(xp^+,\bv'_\mu;\lambda_1)\,\mu^+((1-x)p^+,-\bv'_\mu\, (1-x)/x;\lambda_2)$ \ :
\beqa\label{gammawfb}
\psi_{+\frac{1}{2}+\frac{1}{2}}^{\gamma,+1}(x,\bv_\mu') & = &  \frac{em}{4\sqrt{2}\pi^{2}} K_0\bigl(\frac{M \: b_\mu'}{1-x}\bigr)\nn\\[2mm]
\psi_{+\frac{1}{2}-\frac{1}{2}}^{\gamma,+1}(x,\bv_\mu') & = & \frac{+ieM}{4\sqrt{2}\pi^{2}}\,x\,e^{+i\phi_{b_\mu'}} K_1\bigl(\frac{M \: b_\mu'}{1-x}\bigr)\nn\\[2mm]
\psi_{-\frac{1}{2}+\frac{1}{2}}^{\gamma,+1}(x,\bv_\mu') & = & -\frac{1-x}{x}\, \psi_{+\frac{1}{2}-\frac{1}{2}}^{\gamma,+1}(x,\bv_\mu') \nn \\[2mm]
\psi_{-\frac{1}{2}-\frac{1}{2}}^{\gamma,+1}(x,\bv_\mu') & = & 0  \hspace{.5cm} {\rm and} \hspace{.5cm} 
\psi^{\gamma,\lambda}_{\lambda_1,\lambda_2}(x,\bv'_\mu) = \big[\psi^{\gamma,-\lambda}_{-\lambda_1,-\lambda_2}(x,\bv'_\mu)\big]^*
\eeqa
where $M^{2}=m^2-x(1-x)p^2$. The wave functions were evaluated using the rules for LF wave functions in momentum space given in \cite{Brodsky:1989pv}, Fourier transformed as in \eq{psikb}.

\break


\begin{thebibliography}{999}
\bibitem{Hand:1963zz}
  L.~N.~Hand, D. G. Miller and R. Wilson,
  Rev.\ Mod.\ Phys.\  {\bf 35}, 335 (1963);\\
%
  G.~G.~Simon, C.~Schmitt, F.~Borkowski and V. H. Walther,
  Nucl.\ Phys.\  {\bf A333}, 381-391 (1980).

\bibitem{Drell:1969km}
  S.~D.~Drell and T.~-M.~Yan,
  Phys.\ Rev.\ Lett.\  {\bf 24}, 181-185 (1970).
  
\bibitem{Soper:1976jc}
  D.~E.~Soper,
  Phys.\ Rev.\  D {\bf 15}, 1141  (1977).

\bibitem{Burkardt:2000za}
  M.~Burkardt,
  Phys.\ Rev.\  D {\bf 62}, 071503  (2000)
  [Erratum-ibid.\  D {\bf 66} (2002) 119903]
  [arXiv:hep-ph/0005108];\\
%
  M.~Burkardt,
  Int.\ J.\ Mod.\ Phys.\  A {\bf 18}, 173 (2003)
  [arXiv:hep-ph/0207047].

\bibitem{Diehl:2002he}
  M.~Diehl,
  Eur.\ Phys.\ J.\  C {\bf 25}, 223 (2002)
  [Erratum-ibid.\  C {\bf 31}, 277 (2003)]
  [arXiv:hep-ph/0205208].

\bibitem{Ralston:2001xs}
  J.~P.~Ralston and B.~Pire,
  Phys.\ Rev.\  D {\bf 66}, 111501 (2002)
  [arXiv:hep-ph/0110075].

\bibitem{Brodsky:1989pv}
  S.~J.~Brodsky and G.~P.~Lepage, SLAC-PUB-4947, published in
  Adv.\ Ser.\ Direct.\ High Energy Phys.\  {\bf 5}, 93-240 (1989);\\
%
  S.~J.~Brodsky and H.~C.~Pauli, SLAC-PUB-5558, published in
  Lect.~Notes~Phys.\  {\bf 396}, 51-121 (1991);\\
%
  S.~J.~Brodsky, H.~-C.~Pauli and S.~S.~Pinsky,
  Phys.\ Rept.\  {\bf 301}, 299-486 (1998).
  [hep-ph/9705477].

\bibitem{Brodsky:2000xy}
  S.~J.~Brodsky, M.~Diehl, D.~S.~Hwang,
  Nucl.\ Phys.\  {\bf B596}, 99-124 (2001).
  [hep-ph/0009254];\\
%
  M.~Diehl, T.~Feldmann, R.~Jakob and P.~Kroll,
  Nucl.\ Phys.\  B {\bf 596}, 33 (2001)
  [Erratum-ibid.\  B {\bf 605}, 647 (2001)]
  [arXiv:hep-ph/0009255].


\bibitem{Brodsky:2002ue}
  S.~J.~Brodsky, P.~Hoyer, N.~Marchal, S. Peign\'e and F. Sannino,
  Phys.\ Rev.\  {\bf D65}, 114025 (2002)
  [hep-ph/0104291].

\bibitem{Miller:2007uy}
  G.~A.~Miller,
  Phys.\ Rev.\ Lett.\  {\bf 99}, 112001 (2007)  [arXiv:0705.2409 [nucl-th]];\\
%
  G.~A.~Miller,
  Phys.\ Rev.\  C {\bf 80}, 045210 (2009)
  [arXiv:0908.1535 [nucl-th]].

\bibitem{Carlson:2007xd}
  C.~E.~Carlson and M.~Vanderhaeghen,
  Phys.\ Rev.\ Lett.\  {\bf 100}, 032004 (2008)
  [arXiv:0710.0835 [hep-ph]];\\
%
  C.~E.~Carlson and M.~Vanderhaeghen,
  Eur.\ Phys.\ J.\  A {\bf 41} (2009) 1
  [arXiv:0807.4537 [hep-ph]].

\bibitem{Tiator:2008kd}
  L.~Tiator and M.~Vanderhaeghen,
  Phys.\ Lett.\  B {\bf 672}, 344 (2009)
  [arXiv:0811.2285 [hep-ph]];\\
%
  L.~Tiator, D.~Drechsel, S.~S.~Kamalov and M.~Vanderhaeghen,
  arXiv:0909.2335 [nucl-th].

\bibitem{Pasquini:2007fw}
  B.~Pasquini, D.~Drechsel, L.~Tiator,
  Eur.\ Phys.\ J.\  {\bf A34}, 387-403 (2007).
  [arXiv:0712.2327 [hep-ph]].

\bibitem{Brodsky:2000ii}
  S.~J.~Brodsky, D.~S.~Hwang, B.~Q.~Ma and I.~Schmidt,
  Nucl.\ Phys.\  B {\bf 593} (2001) 311
  [arXiv:hep-th/0003082].
  
\bibitem{Hoyer:2009sg}
  P.~Hoyer and S.~Kurki,
  Phys.\ Rev.\  {\bf D81}, 013002 (2010).
  [arXiv:0911.3011 [hep-ph]].

\bibitem{Lepage:1980fj}
  G.~P.~Lepage and S.~J.~Brodsky,
  Phys.\ Rev.\  {\bf D22}, 2157 (1980).

\bibitem{Isgur:1984jm}
  N.~Isgur and C.~H.~Llewellyn Smith,
  Phys.\ Rev.\ Lett.\  {\bf 52}, 1080 (1984).

\bibitem{Radyushkin:1984wq}
  A.~V.~Radyushkin,
  Acta Phys.\ Polon.\  B {\bf 15}, 403 (1984).

\bibitem{Aitala:2000hb}
  E.~M.~Aitala {\it et al.} [ E791 Collaboration ],
  Phys.\ Rev.\ Lett.\  {\bf 86}, 4768-4772 (2001).
  [hep-ex/0010043];\\
%
  E.~M.~Aitala {\it et al.} [ E791 Collaboration ],
  Phys.\ Rev.\ Lett.\  {\bf 86}, 4773-4777 (2001).
  [hep-ex/0010044].

\bibitem{Brodsky:1973kr}
  S.~J.~Brodsky and G.~R.~Farrar,
  Phys.\ Rev.\ Lett.\  {\bf 31}, 1153 (1973).

\bibitem{Matveev:1973ra}
  V.~A.~Matveev, R.~M.~Muradian and A.~N.~Tavkhelidze,
  Lett.\ Nuovo Cim.\  {\bf 7}, 719-723 (1973).

\bibitem{Anderson:1976ph}
  R.~L.~Anderson {\it et al.},
  Phys.\ Rev.\  D {\bf 14}, 679 (1976).
  
\bibitem{McCracken:2009ra}
  M.~E.~McCracken {\it et al.}  [CLAS Collaboration],
  Phys.\ Rev.\  C {\bf 81}, 025201 (2010)
  [arXiv:0912.4274 [nucl-ex]];\\
%
  R.~A.~Schumacher and M.~M.~Sargsian,
  arXiv:1012.2126 [hep-ph].
  
\bibitem{Bochna:1998ca}
  C.~Bochna {\it et al.}  [E89-012 Collaboration],
  Phys.\ Rev.\ Lett.\  {\bf 81}, 4576 (1998)
  [arXiv:nucl-ex/9808001].

\bibitem{Pomerantz:2009sb}
  I.~Pomerantz {\it et al.}  [JLab Hall A Collaboration],
  Phys.\ Lett.\  B {\bf 684}, 106 (2010)
  [arXiv:0908.2968 [nucl-ex]].

\end{thebibliography}
\end{document}